# THE CLIMATE OF BULGARIA DURING 19<sup>th</sup> and 20<sup>th</sup> CENTURIES BY INSTRUMENTAL AND INDIRECT DATA : SOLAR MODULATED CYCLES AND THEIR EVOLUTION

**Boris Komitov\*, Momchil Dechev \*\* and Peter Duchlev\*\***
*Bulgarian Academy of Sciences -Institute of Astronomy*

*\*- 6003 Stara Zagora-3, Pbox 39 ; b_komitov@sz.inetg.bg ;
\*\* -1784 Sofia-1784, boul. Tsarigradsko shosse 72; mdechev@gmail.bg ;
duchlev@astro.bas.bg*

**Two types of data sets for investigation of solar- modulated cycles in the climate of Bulgaria during the last ~ 200 years has been used in this study: 1. Instrumental data for the rainfalls and temperatures in 26 stations during the period AD 1899-1994; A smoothing dendrochronological data series of the tree rings width of a beech sample (*Fagus*) for the period of AD 1780-1982. The data proceedings has been provided separately for the "winter" (November-April) and "summer" (May-October) half-years. A well expressed quasi-20-22 and 54 yr cycles in the rains and temperature instrumental "summer" series as well as quasi 11yr cycle for the "winter" temeperature data has been established for the studied period 1899-1994. However there are also very serious variations of the 11 and 22 yr cycle magnitudes. The quasi 20-22 yr ("summer") cycle is weak expressed before AD 1930, while the 11 yr (winter) cycle is faded rapidly after ~AD 1975/76. The existence of 20-22 yr climatic cycle could be traced since the end of the supercenturial solar Dalton minimum (AD 1795-1830) in the beech tree ring width series. Strong cycles by duration of 54, 67 and 115 years are also found in this series. All they have solar analogs (in the coronal activity as well as in the "cosmogenic" $^{10}$Be and aurora activity data series). Extreme little tree ring widths during the Dalton minimum has been detected.**

## 1. Introduction

At the beginning of 1980$^{st}$ a very detailed studies of the Bulgarian climate changes during the 20$^{th}$ century has been provided (Komitov, 1986 a,b). The last one has been on the base of a time series analysis of 73 meteorological stations for interval of AD 1899-1979. The following conclusions has been made:

1. On ~3/4 of South Bulgaria territory and separate stations in North Bulgaria a well expressed 20-22 year climatic cycle in the "warm" half- year /May-October/ during the investigated period exist. Most probably it is caused by a solar 20-22 year magnetic /Hale/cycle influence. The cases of the most warmer and dryer summers are predominantly centered to ascedent phases of the even quasi-11year Schwabe-Wolfs solar cycles.
2. A statistical significant quasi-11 year winter temperature cycle on the almost all Bulgarian territory for investigated period has been detected. The most colder winter cases are centered predominantly near to the solar Schwabe-Wolf's cycles minimums
3. There are some evidences about evolution of the solar modulated climatic oscillations during the 20$^{th}$ century.

In this work we give an evidence that the climatic phenomenas during last two decades of 20$^{th}$ century are caused rather by solar activity variations effects as by any other factor. The new study is based on meteorological instrumental data for the period AD 1980-1994 in addition to the previous data sets as well as on indirect climatic data /tree rings width data series from AD 1780 to AD 1982/ and new solar activity data set, based on the Group sunspot annual number *Rh* (Hoyt and Schatten, 1998) ). New statistical methods for cyclic oscillations evolution analysis are used in this paper too. A part of the obtained results and their analysis are presented there. The other ones will be an object of other papers.

## 2. The Data and Methods

For the aims of present study following types of data are used:

1. Instrumental data of mean monthly temperatures and monthly rain sums in 26 meteorological stations for the period 1980-1994 On the base of last ones the mean temperatures and rain sums for "winter"(cold) and "summer"(warm) half- years has been build. The are added to the corresponding data series before 1980 [1];

2. Tree ring width data series / a beech sample (*Fagus*), site near Gurkovo , District of Stara Zagora, Stara Planina Mountain/ for period of 1780-1983-as an indicator for climate changes in Central Bulgaria during the last ~200 years ;

3. The Group Sunspot number annual data series *Rh* for the corresponding time interval , concerning the study , i.e. AD 1780-1994.

For time series analysis and cycles evolution two modifications of the numerical procedure labeled as "T-R periodogram analysis" has been used. The first one is the "standard" T-R periodogram analysis (Komitov, 1986 a,b; 1997). On its base a "mean" spectra of oscillations for the all time series could be obtained. The second one is a "two-dimensional" T-R periodogram analysis /the MWTRPP – *Moving Window T-R Periodogram Procedure* / and it is used for cycles evolution study. The last one is described in details by Bonev et al. (2004) . All above mentioned procedures are placed in 6D-STATv.7.0/7.05/7.1 software package (see: www.astro.bas.bg/~komitov/6d-stat.htm).

## 3 The Results and Analysis

### 3.1 The instrumental climatic data (AD 1899-1994)

As it has been already pointed in &2 we have obtain instrumental data in interval AD 1980-1994 for 26 from the all 73 stations which have been used in the previous study (Komitov, 1986a). Unfortunately there are not published data for the recent years after AD 1994.

On fig.1 the smoothing 5-year average values of warm half- year rain sums in Plovdiv (epoch AD 1899-1994) is shown. On fig 2 the corresponding T-R corellogram is presented. For the last one the following parameters are choice: initial period To=2 years, scan step ΔT= 0.25 years and the total number of scanning steps is 400. As a result the corellogram show all statistical significant cycles in range of 2 to 102 years. As it is visible enough, the 20-22 year cycle remain stable after AD 1980 too. The summer dry period starting near to AD 1984 is in good coincidence with the transition epoch from the odd Schwabe-Wolf's cycle with Zurich number 21st to the even 22nd one. This is in very good agreement with our earlier conclusions (Komitov,1986a) . However it need also note that the dry period after 1984 is the deepest relatively all similar epochs during the instrumental measurements in Bulgaria since AD 1899. On other hand the corresponding summer warming is no so expressive .

The summer temperature dynamics for Plovdiv is shown on fig.3. The 20-22 yr oscillations in opposition to the rain sums ones are well visible.

A new feature on fig.2 is the weak peak near the period T=54 years. By high level of certainty the last one is solar- modulated . A cycle with such duration present in some solar and geophysical data such as the "Greenland" $^{10}$Be data series (Komitov et al, 2003). Because of the fact that this cycle is relatively long it has been not an object of our previous studies in 1980st .

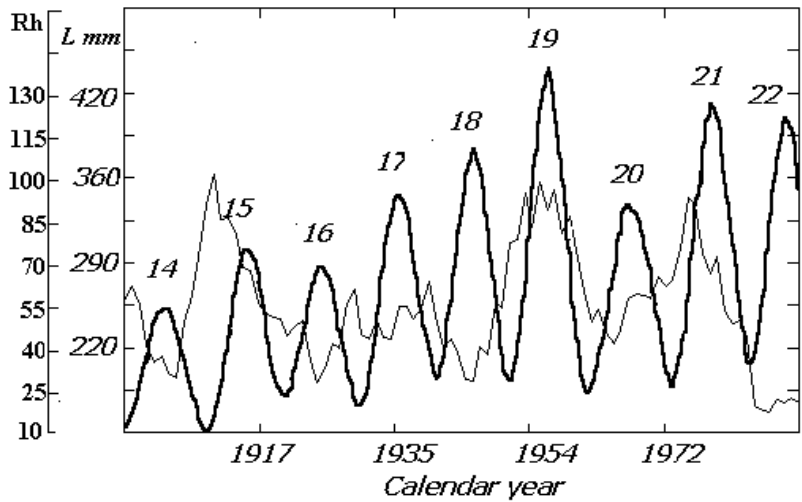

Fig 1. Plovdiv (AD 1899-1994): Smoothing 5-year warm half- year rain sums (thin line) and Group sunspot number index "Rh"(thick line)

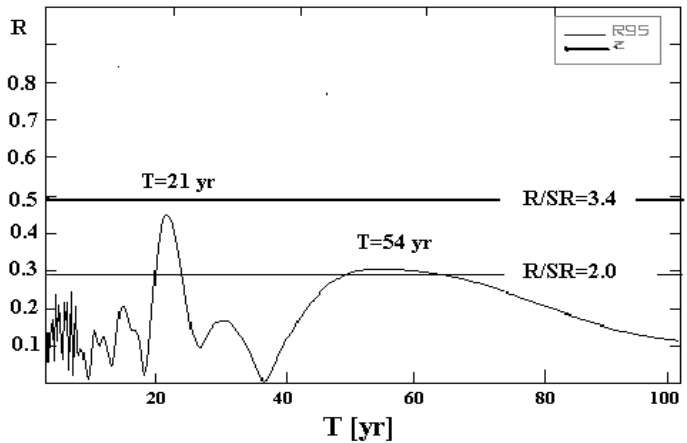

Fig.2. Plovdiv (AD 1899-1994) Warm half- year rain sums : T-R corellogram

On the base of this facts it may to conclude that there are not evidences for essentially different behavior of warm half- year climate in South Bulgaria during AD 1980-1994 as in the previous investigated epoch (AD 1899-1979)(Komitov, 1986a).

As we already are pointed in &1 in the region of West Valleys the quasi-11 year winter temperatures oscillations are best expressed during epoch 1899-1979 (Komitov, 1986a). After an adding of data for epoch 1980-1994 a new analysis has been made.

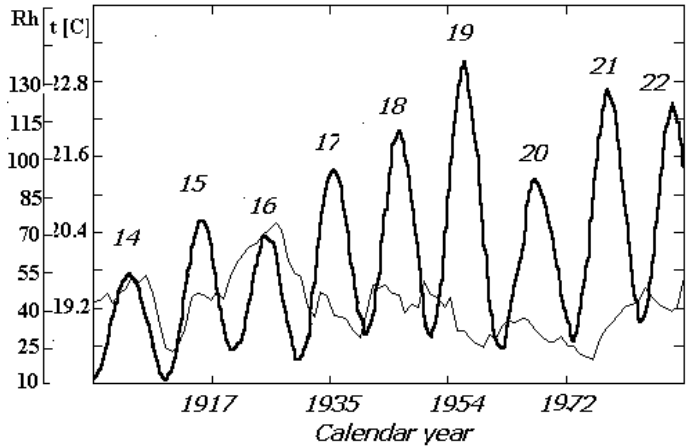

*Fig 3. Plovdiv (AD 1899-1994) Smoothing 5-year warm half- year mean temperatures (thin line) and Group sunspot number index "Rh"(thick line)*

The smoothing 5-years cold half- year mean temperatures for a typical West Valleys station – Dupnitsa ( epoch AD 1909-1994) is shown on fig.4. It is visible that during the last three decades of the investigated period the quasi-11 year oscillation is going down and it is almost totally absent after AD 1980. Unlike the warm half- year most probably there is an influence of some additional factor. On other hand there is no certain general trend in this series and the possibility that the so mentioned 11-year oscillation deceasing is a result of the "antropogenic warming " by our opinion must to exclude. An possible alternative cause may connected to large scale solar activity regime changes. As it is shown in an our new study a very important cause could be the long term change of the north- south sunspot area asymmetry sign from positive to negative, which has been occur near to AD 1975/76 (Komitov, 2010).

To test the hypothesis for large-scale solar activity regime influence over the climate changes we use two methods in this study. The first one is by using a T-R periodogram "moving epoch " procedure , i.e. "two-dimensional" T-R periodogram analysis [5]. It is based on the "standard" T-R procedure, which is used not over the all series, but in small its parts with equivalent length. Every one part is shifted relatively to the adjacent ones by equidistant time step $\Delta t$ ("moving window"). In our case this time step of shifting is equal to 1 year. By this method we derive from the general time series a large number of shorter parts. Over every one such "moving epoch" a standard T-R periodogram analysis with constant choiced parameters To, $\Delta T$ and scanning steps number is proceeded. As a result a map , presented coefficients of correlation between F(t) and $\varphi(t)$ is derived. F(t) are the original ,observed data values and $\varphi(t)$ is a simple periodic function with current period T, obtained by a mean least squares procedure (Komitov, 1986,a,b; 1997). Every column in this map correspond to a single standard T-R corellogramm for a separated "moving epoch". The last one is presented by number of column. The coefficients of correlation values R are presented on map by colors or gray halftones scale. A more expressive variant is to show map no of the R values , but of corresponding R/SR ,where SR is the error of R.

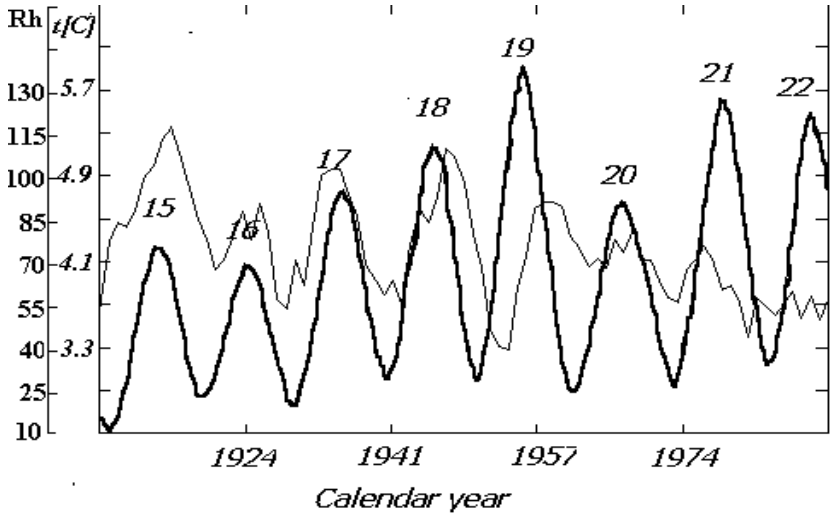

*Fig 4. Dupnitsa (AD 1909-1994) Smoothing 5-year winter half- year mean temperatures (thin line) and Group sunspot number index "Rg"(thick line)*

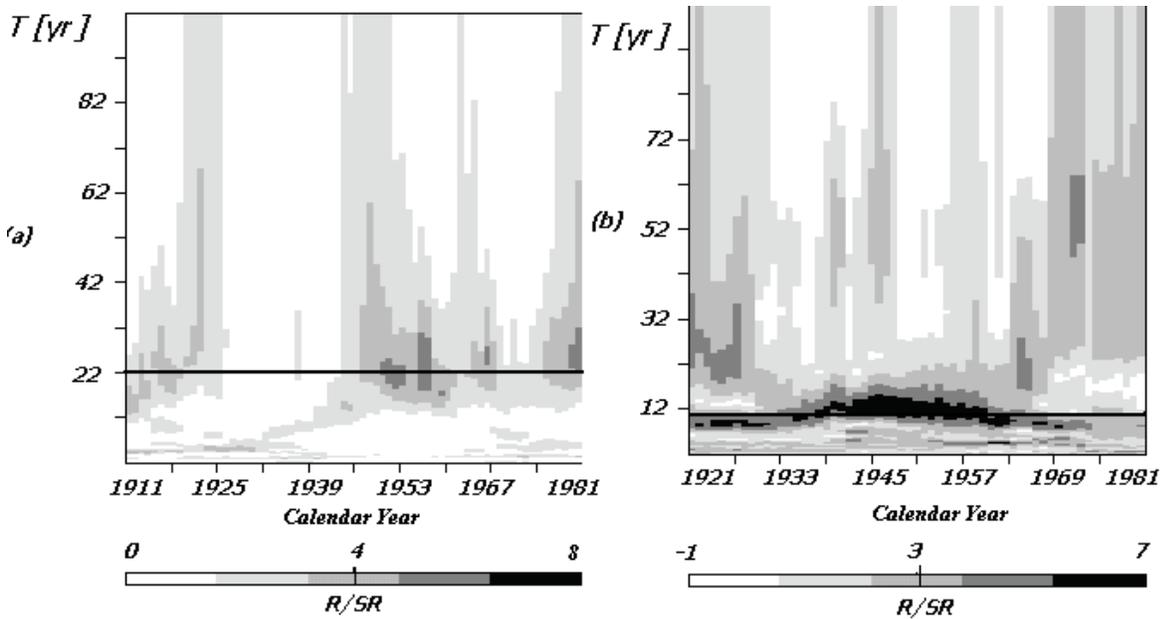

*Fig 5 a,b  The two-dimensional T-R corellograms; left -Plovdiv (1899-1994)    - rain sums ,warm half- year; right –Dupnitsa (1909-1994) –mean temperatures, cold half-year. On the 'X'-axis the centers of the moving epochs are presented*

For the aims of the present study we use a standard "moving window" length of 25 years and the following parameters of "standard " T-R corellogram: To=2 years, ΔT=0.25 years and 100 scanning steps.

The results for Plovdiv (the warm half- year rain sums) and Dupnitsa (mean cold half- year temperatures) are shown on fig 5a and 5b respectively. It may conclude, that the amplitudes and significance of quasi 11 and 20-22 years cycles (presented by R/SR ratio) are varied essentially during the 20$^{th}$ century. It may saying that generally at the beginning of the century, which approximately coincided with the start of the last centurial solar cycle the both solar modulated climatic cycles are relatively weaker as in the middle of century, corresponding of centurial as well as of quasi- bicenturial solar cycles maximums [7]. After ~AD 1960 (i.e. after Schwabe-Wolf's cycle No 19 maximum) the amplitude of quasi 11-year climatic oscillation is going down. The absolute minimum of the 20-22 year cycle magnitude is near to AD 1930. During the last 2-3 decades of century a new increasing of 20-22 years oscillation in warm halfeyar is observed.

The above mentioned results are indicated that the shorter (11 and 22 years) solar-climatic cycles amplitudes are depended by large time scale solar-climatic relationships tendencies.

*3.2 Solar-Modulated Cycles in the Climate of Central Bulgaria (AD 1780-1982) by dendrochronological data*

In this paragraph we give brief comments of the main results concerning the data analysis of a beech sample tree rings widths. The age of the last one is 209 years. The initial six years are removed from investigated time series because of the high level uncertainty . The final series is related to the epoch of AD 1780-1982. The measurement precision is ~0.05 mm. The first results and analysis have been presented in 2001 (Komitov and Vladimirov, 2001).

The smoothing 5-years tree rings widths are shown on fig.6. The corresponding T-R corellogram is on fig.7. The quasi-20 year oscillations with various amplitude are presented in almost all time series, except at the beginning of 19$^{th}$ century. The last one period is characterized by very small values of tree rings width (~0.1-0.15 mm). It is an interesting epoch for large scale variations of solar activity – the so called "Dalton minimum". The last one is a supercenturial solar minimum , coincided with the declining phase of solar quasi- bicenturial (~205 year) cycle.

In the T-R spectra (fig 7) the cycles with duration of 54, 67 and 115 as well as a 'multiplet' in the range 18-24 years are statistically significant. It need note that oscillations by duration of 54 , 67 and 115-120 years are detected in $^{10}$Be Greenland series ( 1423-1985) as well as in the middle latitude aurora activity (Komitov et al., 2003; Komitov,2009). The last ones (67 and 115-120 years) are very strong in this series. A 115-120 year cycle has been found by Duchlev (2002) as a quasi- periodic trend -"hyper- cycle" in the Meudon time series of the quiet filaments. These facts are indicated that in corona and solar wind activity are very important also some processes ,where are with essentially different nature and time behavior as these ones in photosphere. Most probably this type of activity is related to the large time scale solar variations and is characterized by high climatic efficiency. It is interesting also to point out that a quasi- 120 yr oscillation is detected in the north-south sunspot area asymmetry (the Pulkovo Observatory archive data from AD 1821 to 1994) (Komitov, 2010)

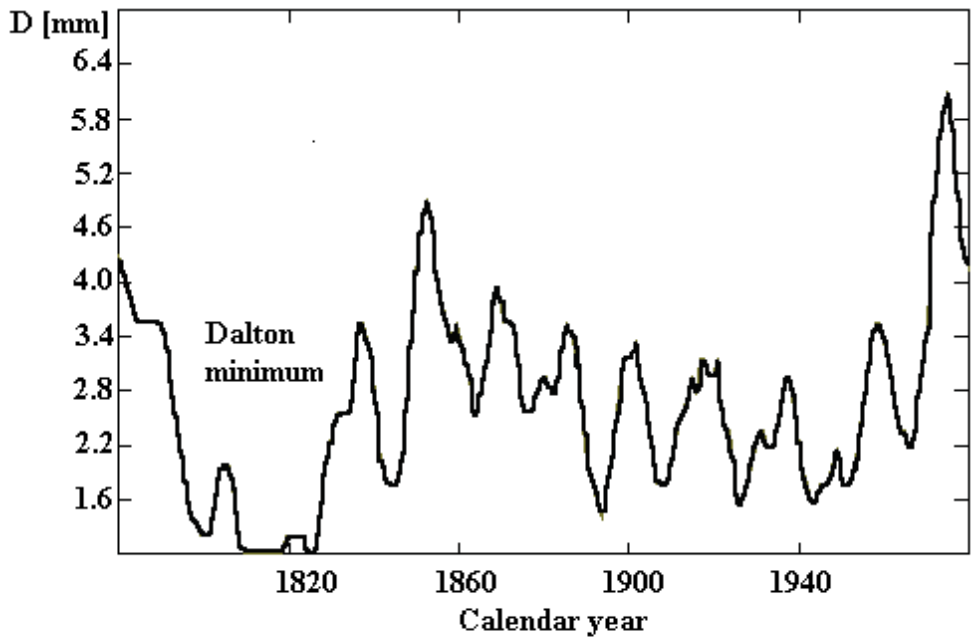

*Fig 6. Beech tree rings width (AD 1780-1982) (Gurkovo, District of Si.Zagora) smoothing 5-year values*

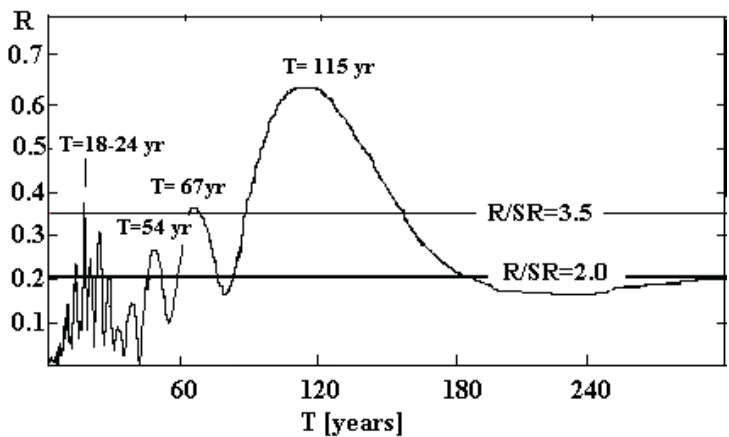

*Fig 7. Beach tree rings width (AD 1780-1982) ;T-R corellogram*

An additional evidence in this course is the fact that 20-22 cycles magnitude is modulated in our beech tree ring width data by 67 year cycle. The minimal 20-22 years oscillation magnitudes are near to Dalton minimum, AD 1870 and 1930-1940 respectively (Komitov and Vladimirov, 2001). The last one is in good coincidence with the 20-22 year warm half- year rain cycle magnitudes behavior by instrumental data (see above).

On the base of all significant cycles , which are shown in the T-R corellogram on fig.7 a time series model of the tree ring width data has been build using technique, described by Komitov(2001) . Its extrapolation after AD 1982 close to 2030 show rapid tendency to for decreasing of tree rings widths to values of ~0.1 mm , i.e. very similar to these during Dalton minimum. The deepest phase (narrowest tree rings) should be occur near to AD 2010. This result is very significant and interesting if it take into account that a coming of supercenturial solar minimum during $21^{st}$ century is predicted by Komitov and Kaftan (2003, 2004). In climatic aspect it may indicated summer drying tendency up to ~ AD 2010-2012 and long term period of summer cooling and high rain levels then.

*4. Conclusions*

On the base of presented in this study results and its analysis the following conclusions can be made:
1. The solar modulated climatic cycles by duration of 10-11 and 20-22 years exist in climate of Bulgaria and it can detected in instrumental as well as in indirect data series. Its magnitude varied by the modulation of longer solar -climatic cycles by subcenturial and quasicenturial duration.
2. The 54, 67 and ~115 year climatic cycles have analogies in solar corona , solar wind and geomagnetic phenomena behavior . They indicated that a significant part of solar influence over Earth climate may be related to processes where are running not in photosphere , but in outer parts of Sun\s atmosphere.